\documentstyle[preprint,aps]{revtex}
\newcommand{\eq}[1]{eq.(\ref{#1})}
\begin{document}
\tighten
\preprint{PSU/TH/186; hep-ph/9707381}
\title{Radiative Correction to the Nuclear-Size Effect and 
Hydrogen-Deuterium Isotopic Shift} 
\author {Michael I. Eides \thanks{E-mail address:  
eides@phys.psu.edu, eides@lnpi.spb.su}}
\address{ Department of Physics, Pennsylvania 
State University, 
University Park, PA 16802, USA\thanks{Temporary address.}\\ 
and
Petersburg Nuclear Physics Institute,
Gatchina, St.Petersburg 188350, Russia\thanks{Permanent address.}}
\author{Howard Grotch\thanks{E-mail address: h1g@psuvm.psu.edu}}
\address{Department of Physics, Pennsylvania State University,
University Park, PA 16802, USA}
\date{July, 1997}

\maketitle
\begin{abstract}
The radiative correction to the nuclear charge radius contribution to the 
Lamb shift of order $\alpha(Z\alpha)^5m_r^3<r^2>$ is calculated. In view of 
the recent high precision experimental data, this theoretical correction 
produces a significant contribution to the hydrogen-deuterium isotopic 
shift.
\end{abstract}

\section{Introduction}

Spectacular experimental results were obtained in recent years for the  
intervals between levels with different principal quantum numbers and for  
the ground state $1S$ Lamb shift in hydrogen and deuterium 
\cite{pach,bhb,bou,bea}. Experimental error for the isotopic shift between 
the hydrogen and deuterium in the $1S-2S$ transition was reduced to $0.2$ 
kHz (see \cite{hanch} and a private communication from H\"ansch, cited 
in \cite{friarmay}).

These experimental achievements have created new challenges for the 
hydrogen-deuterium isotopic shift theory, which now should account at least 
for all corrections to the order of one tenth of kHz. In a surge of 
theoretical activity there appeared a number of papers 
\cite{pach93,pach94,luros,leiros,msz,milkhr,friarpayne,friarapr,friarmay,khrsen,friarjul}, 
where corrections to  the isotopic shift induced by the nuclear charge 
radius and polarizablity were considered, and a set of old results in the 
field \cite{berner,friar79} were either rederived or improved. However, to 
the best of our knowledge the radiative correction of order 
$\alpha(Z\alpha)^5m_r^3<r^2>$ to the main nuclear radius contribution was 
not taken into account in discussion of the recent experimental data. As we 
will show below this correction is about $0.6$ kHz for the 
hydrogen-deuterium isotopic shift and should be considered on par with the 
other contributions.

\section{Radiative Correction to the Finite Size Effect}
								   
The radiative correction to the finite size effect was first discussed in 
\cite{borie}, where a very large contribution was obtained.  The situation 
was almost immediately clarified in \cite{lye}, where it was shown that the
respective contribution is generated by the region with large intermediate 
momenta and should actually be a small correction of order 
$\alpha(Z\alpha)^5m_r^3<r^2>$. Relying on the estimate of \cite{lye} the 
authors of \cite{sy} anticipated a contribution of about 10 Hz for the $2S$ 
state in hydrogen. The error in \cite{borie} was connected with an 
erroneous extrapolation of a nonrelativistic approximation to the 
relativistic momenta. This led to an overestimate of the resulting 
contribution. 

Let us consider briefly calculation of the leading nuclear size contribution 
to the energy shift. It is generated by the slope of the nuclear formfactor 
in its low-momentum expansion

\begin{equation}   \label{formfactor}
F({k}^2)\approx 1-\frac{k^2}{6}<r^2>,
\end{equation}

where the dimensionless momentum $k=|{\bf k}|$ is measured in the units of 
the electron mass. The momentum squared in the second term above cancels the 
$1/k^2$ in the Coulomb photon propagator and leads to a momentum 
independent perturbation potential 

\begin{equation}
\Delta V=\frac{2\pi(Z\alpha)}{3}<r^2>.
\end{equation}

Then we immediately obtain

\begin{equation}
\Delta E=\frac{2\pi(Z\alpha)}{3}<r^2>|\psi(0)|^2
=\frac{2(Z\alpha)^4}{3n^3}m_r^3<r^2>.
\end{equation}

Any radiative correction behaves as ${k}^2$ at small exchanged 
momenta, and the presence of such a correction pushes the significant 
integration momenta to the relativistic region for the electron. The 
skeleton integrand approach (see, e.g., \cite{eksann1,ego}) is ideally 
suited for calculation of such corrections.

The calculation essentially coincides with the calculation of 
corrections of order $\alpha^2(Z\alpha)^5$ to the Lamb shift in the 
skeleton integral framework \cite{ego,eg,eg4,eg5,esjetp,es} but is 
technically simpler due to the simplicity of the nuclear formfactor slope 
contribution in \eq{formfactor}. The skeleton contribution to the Lamb 
shift, induced by the diagrams with two external photons, has the form 
\cite{ego}

\begin{equation}            \label{skel}                 
\Delta E_{\mbox{skel}}=-\frac{16(Z\alpha)^5}{\pi 
n^3}(\frac{m_r}{m})^3m\int_0^\infty dk\frac{R(k)}{{k}^2}, 
\end{equation}

where $R(k)$ is the factor describing the radiative and the 
nuclear structure insertions.

There are two kinds of radiative insertions: 
one-loop polarization insertion in one of the external Coulomb lines, and 
one-loop radiative insertions in the electron line. It is clear that the 
magnitude of the radiative corrections to the nuclear size contribution 
grows with the nuclear charge as $Z^5$, and their importance increases for 
the highly charged ions. The respective results for the highly charged ions 
are well known (see, e.g., \cite{mohrsoff} and references therein).  
The radiative correction to the nuclear size effect in hydrogen induced by 
the polarization operator insertion was calculated in \cite{hylton}, and to 
the best of our knowledge the respective correction induced by the radiative 
insertions in the electron line was never discussed in the literature. 
We will now calculate both of these corrections. 

\subsection{Polarization Correction}

For calculation of the polarization correction we have to insert in the 
integral in \eq{skel}  the polarization operator 

\begin{equation}
\frac{\alpha}{\pi}{I_1(k)}= \int_0^1 dv \frac{v^2(1-v^2/3)}{4+(1-v^2)k^2}\:,
\end{equation}

instead of the function $R(k)$, and also insert in the integrand 
the nuclear slope contribution from \eq{formfactor}. Then the respective 
contribution to the energy shift has the form

\begin{equation}
\Delta E_{\mbox{pol}}=\frac{32\alpha(Z\alpha)^5<r^2>m^2}{3\pi^2 
n^3}(\frac{m_r}{m})^3m\int_0^\infty dkI_1(k), 
\end{equation}

where we have inserted an additional factor 4 in the integral in order to 
take into account all possible ways to insert the polarization operator and 
the slope of the nuclear formfactor in the Coulomb photons. 

After an easy analytic calculation we obtain, in complete agreement with 
\cite{hylton},

\begin{equation}                       \label{pol}
\Delta E_{\mbox{pol}}=\frac{1}{2}m_r^3<r^2>\frac{\alpha(Z\alpha)^5}{n^3}.
\end{equation}

\subsection{Electron-Line Correction}

For calculation of the electron-line correction we have to insert in the 
integral in \eq{skel}  the electron line factor $L(k)$ \cite{eg4,egs} 
instead of the function $R(k)$, and also insert in the integrand 
the nuclear slope contribution from \eq{formfactor}. Then the respective 
contribution to the energy shift has the form

\begin{equation}
\Delta E_{\mbox{e-line}}=\frac{16\alpha(Z\alpha)^5<r^2>m^2}{3\pi^2 
n^3}(\frac{m_r}{m})^3m\int_0^\infty dkL(k), 
\end{equation}

where we have inserted an additional factor 2 in the integral in order to 
take into account all possible ways to insert the slope of the nuclear 
formfactor in the Coulomb photons. 

After numerical calculation we obtain

\begin{equation}          \label{eline}
\Delta E_{\mbox{e-line}}=-1.985(1)m_r^3<r^2>\frac{\alpha(Z\alpha)^5}{n^3}.
\end{equation}

In principle, this integral also admits an analytic evaluation in the same 
way as it was done for a more complicated integral in \cite{egs}.

\section{Discussion of Results}

The total radiative correction to the nuclear size effect is given by the 
sum of contributions in \eq{pol} and \eq{eline}

\begin{equation}          \label{tot}
\Delta E=-1.485(1)m_r^3<r^2>\frac{\alpha(Z\alpha)^5}{n^3}.
\end{equation}

Numerically the contribution to the hydrogen-deuterium isotopic shift for 
the $1S-2S$ interval is equal to

\begin{equation}
\Delta E(1S-2S)_{D-H}=-0.616\; \mbox{kHz},
\end{equation}

where we have used $<r^2>^{1/2}_D=2.128(11)$ fm \cite{sitr} for the 
deuteron radius, and $<r^2>^\frac{1}{2}_H=0.862(12)$ fm \cite{ssbw} for the 
proton radius.

This correction is clearly phenomenologically relevant on the background of 
the experimental uncertainty which is now about $0.2$ kHz for the isotopic 
shift.

\acknowledgements
M. E. is deeply grateful for the kind hospitality of the Physics 
Department at Penn State University, where this work was performed. The 
authors appreciate the support of this work by the National Science 
Foundation under grant number PHY-9421408.

\bigskip
{\it Notes Added in Proof.}
The vacuum polarization correction to the finite size 
effect was considered earlier by J. L. Friar, Z. Physik, A 292, 1 (1979);
(E) A 303, 84 (1981). For the spherical distribution of charge the result of 
this calculation coincides with our Eq.(7) and the result of Ref.[30].  We 
are grateful to  J. L. Friar for bringing his work to our attention.

The radiative correction to the nuclear size effect was considered earlier 
by K.  Pachucki, Phys. Rev. {\bf A48}, 120 (1993) as a radiative correction 
to the electron charge density. Restoring the missing factor $\pi$ in 
Eq.(88) in this work (an apparent misprint) one obtains from Eq.(85) there 
the value 1.431 instead of the numerical coefficient 1.485 in our Eq.(10) 
above. In the calculations of K. Pachucki the contributions of the vacuum 
polarization and the electron factor were not separated, so we were unable 
to find out the source of the 0.05 discrepancy in the value of the 
coefficient in Eq.(10).  We are grateful to S. Karshenboim who attracted our 
attention to this work by K. Pachucki.

\end{document}